\documentclass[%
 reprint,
 superscriptaddress,
 onecolumn,
 amsmath,amssymb,
 aps,
pra,
]{revtex4-1}

\usepackage[paperwidth=210mm,paperheight=297mm,centering,hmargin=2.3cm,tmargin=2.8cm,bmargin=3.4cm]{geometry}

\usepackage{graphicx}
\usepackage{amsfonts}
\usepackage{amssymb}
\usepackage{amsmath}
\usepackage{xcolor}
\usepackage{float}

\newcommand{\be}{\begin{equation}}
\newcommand{\ee}{\end{equation}}
\newcommand{\bea}{\begin{eqnarray}}
\newcommand{\eea}{\end{eqnarray}}

\def\lb{\label}

\newcount\bozza \bozza=0
\ifnum\bozza=1
\newdimen\shift \shift=-2truecm
\def\lb#1{%
{\label{#1}\rlap{\kern\shift{$\scriptstyle#1$}}}}
\else\def\lb#1{\label{#1}} \fi

\begin{document}

\title{Effect of inter-well interactions on non-linear beam splitters for matter-wave interferometers}

\author{C. Baroni}
\
\affiliation{Institut f\"ur Quantenoptik und Quanteninformation (IQOQI),
  \"Osterreichische Akademie der Wissenschaften, 6020 Innsbruck, Austria}
 \affiliation{Institut f\"ur Experimentalphysik, Universit\"at Innsbruck, Austria}

\author{G. Gori}
\affiliation{Institut f\"ur Theoretische Physik, Universit\"at 
  Heidelberg, D-69120 Heidelberg, Germany}

\author{M. L. Chiofalo}
\affiliation{Dipartimento di Fisica ``Enrico Fermi'', Universit\`{a} di
  Pisa and INFN, Largo B. Pontecorvo 3, I-56127 Pisa, Italy}

\author{A. Trombettoni}
\affiliation{Department of Physics, University of Trieste, Strada Costiera 11, I-34151 Trieste, Italy}
\affiliation{CNR-IOM DEMOCRITOS Simulation Center and SISSA, Via Bonomea 265, I-34136 Trieste, Italy}

\begin{abstract}
We study the non-linear beam splitter in matter-wave interferometers using ultracold quantum gases in a double-well configuration in presence of non-local interactions inducing 
inter-well density-density coupling, as they can be realized, e.g., with dipolar gases. 
We explore this effect after considering different input states, in the form of either coherent, or Twin-Fock, or NOON states.
We first review the non-interacting limit and the case in which only the local interaction is present, including the study
of sensitivity near the self-trapping threshold. Then, we consider the two-mode model in the presence of inter-well interactions and consider the scaling of the sensitivity as a function of the non-local coupling strength. Our analysis clearly shows that non-local interactions can compensate the degradation of the sensitivity induced by local interactions, so that they may be used to restore optimal sensitivity.
\end{abstract}
\maketitle

\section{Introduction}
Quantum metrology is raising as a timely research field, where suitably engineered correlations among many particles represent a new paradigm to push sensitivity to unprecedented limits. For example, it is well known that the sensitivity of atom interferometers such devices can be improved by using squeezed states as input states, or taking advantage of interactions in the splitting process \cite{pezze}. However, in experimental setups, interactions may lead to a degradation of the sensitivity \cite{PS06}, so that a general issue is to explore how the performance of quantum interferometers depends
on the parameters and the features of the interatomic couplings. In this paper, we study the effect of non-local interactions among
particles on the sensitivity of a matter-wave interferometer realized with ultracold atoms \cite{molecules,Gross_2012,pezze}.

The remarkable features of ultracold atoms \cite{BDZ}
make them an ideal platform in which
quantum interferometers can be implemented, with the perspective of making matter-wave interferometers excellent tools for precision measurements
of several physical quantities, such as accelerations and rotations
\cite{segnac}, and to test basic quantum-mechanics \cite{QMCP}
and general relativity \cite{Newton} concepts. To this aim, an especially suited setup is provided by
quantum gases in a double-well potential, with the possibility to tune the height of the barrier and the energy difference between the two wells \cite{Smerzi97bis}. A similar scheme can be realized after using Rabi-coupled Bose gases \cite{Williams99}. Both configurations give rise to what is also referred to as
the external (in physical space) and internal (in the atomic species) ultracold Josephson effect, and are well described by two-mode (2M) models in the regime in which the two gases
are weakly coupled. These two configurations have both been experimentally investigated also in the case
of multi-well potentials, and for bosonic and fermionic ultracold gases \cite{Cataliotti843,Jos_osc,Smerzi2003,Primo_inter,on_chip,Anker,Levy2007,Hall2007,Esteve2008,NL_Gross,ACB,LeBlanc,twin,LENS,QPT,Fattori2017,Burchianti18,Xhani20}.

In essence, a general (matter-wave) interferometer is an instrument to measure the effect of a given physical process acting in a differential manner on an input wave. In fact, the input wave passes through a first suited beam splitter and is separated in two paths, where the two evolving waves accumulate a phase difference driven by the given physical process to be tested. In order to enhance the visibility of the effect, a possibility is, e.g., that the paths run as far as possible away from each other. A second beam splitter then recombines the two paths, so that  the accumulated phase difference can be inferred from the readout.
In the double-well setup for a matter-wave interferometer, the splitting process is realized by letting the barrier being high enough, in order to accumulate a phase difference between the atoms in the two wells \cite{pezze}.
After a phase difference is accumulated, recombination is performed by lowering the barrier back or by letting the atom clouds fall and overlap during their expansion.

The presence of interactions among the atoms in a matter-wave interferometer leads to several relevant effects that have been investigated in the literature ~\cite{SSN,Phase_Sens,PS06,demler,Entang_Squeez,Mach-Smerzi,sinatra,Wrubel2018,Zhang2019}. In particular, these include
the connection between the Heisenberg limit, in which sub-shot noise can be achieved,
and multiparticle entanglement useful for metrology, measured by the Fisher information \cite{Entang_Squeez,pezze,Gross_2012,RMP}.
Given the essential role that the beam splitter plays in interferometric scheme, it is crucial to study the interaction effects occurring during the beam-splitting process and the corresponding non-linear contributions to the scaling of the phase sensitivity
with the total number of particles. These effects have been extensively investigated in the literature \cite{PS06,Huang08,Grond09,Jin,Mazzarella_2011,Chwede,Mullin,Prakash,Recomb} (see also  \cite{Gross_2012} and references therein).
In \cite{PS06}, three regimes, namely Rabi, Josephson and Fock, were
distinguished in the presence of local interactions, associating to them different scalings of the phase
sensitivity with the total number of particles.

The non-local nature of interparticle interactions represents one additional issue to be considered in atomic, molecular and optical systems, where 
  several studies have been carried out on how equilibrium properties and quantum dynamics are affected and the long-rangeness of the interactions possibly exploited ~\cite{vodola14,gong16,lepori16,defenu16,blass18,defenu18,lerose19}. In cold atoms, these characterize for example the physics of dipolar gases \cite{Lahaye_2009}. In the context of ultracold quantum interferometers, the natural question arises on how their performance can be modified. In the present paper, we provide the first segment of the answer to this question, by focusing our attention on the physics of the non-linear beam splitters in the presence of non-local interactions.
We find that that non-local interactions can compensate the degradation of the sensitivity induced by local interactions, which can therefore be engineered to restore optimal sensitivity.

The paper is organized according the following scheme.
In Section \ref{themodel} we introduce the 2M model, we present a brief discussion of the self-trapped regime occurring in a double-well above a critical population imbalance and then remind basic properties of ultracold quantum interferometers. In section \ref{inter-well section} we describe the effect of inter-well interactions. 
The analysis of the ultracold beam splitter in the presence of non-local interactions is discussed for different input states in Section \ref{BS}. 
Our conclusions are presented in Section \ref{conclusions}.
 
\section{The Model}\label{themodel}

In the 2M approximation, the many-body description of $N$ bosons in a double-well potential interacting via local interparticle interactions
can be simplified in the form of a two-sites Bose-Hubbard model with the following Hamiltonian:
\begin{equation}
 \hat{H}= -J(\hat {a}^\dagger \hat b+ \hat {b}^\dagger \hat a) + \frac{U}{2}(\hat{a}^\dagger\hat{a}^\dagger \hat a \hat a + \hat{b}^\dagger\hat{b}^\dagger \hat b \hat b ) + \frac{\delta}{2}(\hat{a}^\dagger\hat{a}-\hat{b}^\dagger\hat{b}) \; ,
\label{H1}
\end{equation}
where $\hat a^\dagger$  $(\hat b^\dagger)$  creates a particle in the left (right) well and $\hat a$ $(\hat b)$ destroys it, 
$J$ is the tunnelling strength, $U$ the interaction strength in each well and $\delta$ the energy shift between the two wells. These quantities can all vary with time: during the interferometric process, $U$, $J$ and $\delta$ need to be varied and are thus time-dependent.
In the following, we denote with  $\hat{n}_a=\hat{a}^\dagger\hat{a}$ and $\hat{n}_b=\hat{b}^\dagger\hat{b}$ the number operators in each well.
The elements of the Fock space are denoted as $|n\rangle$,
with $|n\rangle \equiv |n_L=n; n_R=N-n\rangle$. 

\begin{figure}[H]
\centering
 \includegraphics[width = 0.55\textwidth]{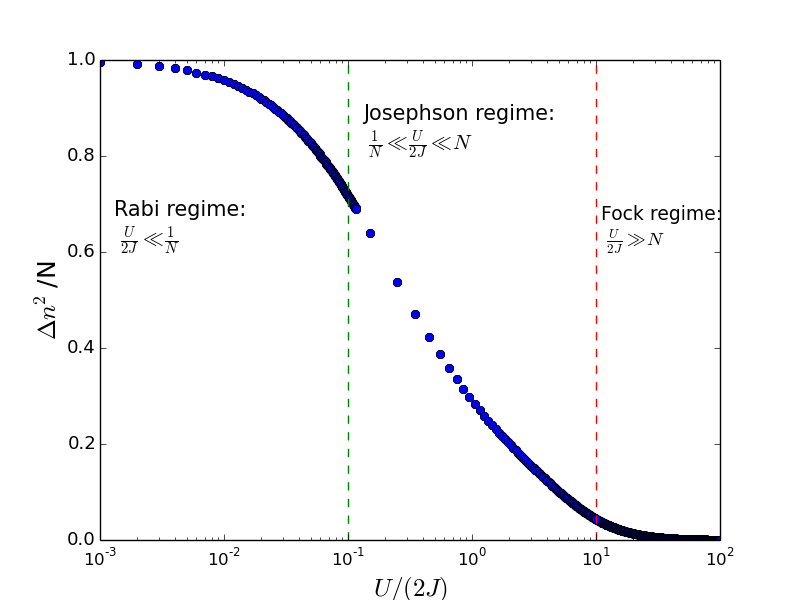}
 \caption{Variance of population imbalance $vs.$ interaction energy $U/(2J)$, characterizing the Rabi, Josephson, and Fock regimes (see text). $N=10$.}
 \label{fig_1}
\end{figure}

While we refer to \cite{Gross_2012,Dalton} for a detailed discussion on the 2M model and its applications to interferometry, we remind readers here of the essential properties. First, the ground state of Hamiltonian (\ref{H1}) can be written in the form $|\Psi\rangle=\sum_{n=0}^N c_n |n\rangle$.
Second, considering a symmetric double-well ($\delta = 0$),
Hamiltonian (\ref{H1}) depends only on the parameter $\gamma = {U}/({2J}) $, and the three regimes in Figure~\ref{fig_1} can be identified \cite{leggett}: 
\begin{itemize}
\item [(i)] {\it Rabi} with $\gamma \ll {1}/{N}$. Here, interactions are negligible and the tunnelling term dominates. The energy spectrum is linear, such as in an harmonic oscillator with levels separated by $2J$;
\item [(ii)] {\it Fock} with $\gamma \gg N$. Here, interactions dominate and the spectrum has a quadratic form, such as in a pairwise quasi-degenerate states with opposite imbalance ($|n_L,n_R\rangle, |n_R,n_L\rangle$); 
\item [(iii)] {\it Josephson} with ${1}/{N} \ll \gamma \ll N$. Here, both interactions and tunnelling play a relevant role. The spectrum starts with a linear behavior and then becomes quadratic with pairwise quasi-degenerate states.
\end{itemize}

In addition, one can see that
\begin{itemize}
\item [(a)] For large and negative $\gamma$, the ground state is the NOON state $({|N,0\rangle + |0,N\rangle})/{\sqrt{2}}$ \cite{noon}.
\item [(b)] For attractive energies such that ${1}/{N} \ll |\gamma| \ll N$, the energy spectrum starts with a quadratic behavior, with pairwise degenerate states and then it becomes linear with levels separated by $2J$.
\item [(c)] For $|\gamma|\gg N$, the energy spectrum is quadratic with pairwise degenerate states for large negative values of the interaction energy.
\end{itemize}
The qualitative behavior of the energy spectrum is well confirmed by numerical results \cite{Tesi_Cosetta}.

Concerning the variance of the population imbalance, two limiting cases can be identified. For $U=0$,
the Hamiltonian is diagonal in the basis of the ground and first excited states: $\hat H \equiv E_g \hat{a}^\dagger_g\hat{a}_g + E_e \hat{a}^\dagger_e\hat{a}_e$, and $J=({E_e-E_g})/{2}$. The ground state corresponds to 
 \begin{equation}
 |\Psi\rangle = \frac{1}{\sqrt{N!}} (\hat{a}^\dagger_L + \hat{a}^\dagger_R)^N|0\rangle = \frac{1}{\sqrt{2^N}} \sum_{n=-N/2 }^{N/2} \frac{N!}{(\frac{N}{2}+n)!(N-\frac{N}{2}-n)!}|n\rangle \; . 
 \end{equation}
 For $J=0$, the Hamiltonian is diagonal in the basis of Fock states $|n_L,n_R\rangle$ (assuming an even number of particles)
 with eigenvalues ${U}(n_L^2+n_R^2-N)/2$.
 The minimum value for $n_L^2+n_R^2$ with the constraint $n_L+n_R=N$ is reached for $n_L=n_R$, so that the ground state is a Twin Fock state:
 \begin{equation}
  |\Psi\rangle=TF=|N/2,N/2\rangle \; .
 \end{equation}
In this state, $\Delta n^2=0$ and the phase is uniformly distributed (random phase).

As shown by the numerical results obtained from the solution of Hamiltonian (\ref{H1}) and displayed in Figure \ref{fig_1}, the variance in the three regimes (i)-(iii) defined above results to be:
 \begin{itemize}
\item [(i)] {\it Fock regime}:  
    $\Delta n = {JN}/(\sqrt{2}{U})\ll 1 $;
\item [(ii)] {\it Rabi regime}:
    the ground state is still close to a coherent state and one has  $\Delta n \lesssim \sqrt{N}/2 $;
 \item [(iii)] {\it Josephson regime}:
    for low-lying energy states one can
    see that $\Delta n^2 = N/(4\sqrt{1+\Lambda})$.
 \end{itemize}
An analytical estimate of the variance is given by \cite{JI99}:
 \begin{equation}
  \Delta n = \sqrt{\frac{N}{2}} \bigg( \frac{2J}{2J+NU} \bigg)^\frac{1}{4} \; .
  \label{variance_JI}
 \end{equation}


\subsection{Self-Trapping Regime}
\label{self}

For a condensate in a double-well potential
it is possible to define the relative phase between the atoms in the two wells $\phi = \phi_L - \phi_R$ and the fractional population imbalance $z = ({n_a -  n_b})/{N}$. In a mean field picture, the equations
of motion for these quantities can be cast in the form
\cite{Smerzi97bis,Smerzi97}:
\begin{equation}
 \dot z(t) = -\sqrt{1-z^2(t)} \sin(\phi(t)) \; ,
  \label{motion1}
\end{equation}
\begin{equation}
 \dot \phi(t) = \Delta E + \frac{UN}{2J}z(t) + \frac{z(t)}{\sqrt{1-z^2(t)}} \cos(\phi(t)) \; ,
  \label{motion2}
\end{equation}
where $\Delta E = \delta/(2J)$.

When the interaction strength is large enough compared to the tunnelling rate,
Equations (\ref{motion1}) and (\ref{motion2}) imply a non-linear behavior
and an exact solution for $z(t)$ can be found in terms of elliptic
functions \cite{Smerzi97bis,Smerzi97}. If the initial population imbalance is non vanishing, the sinusoidal oscillations around $z = 0$ become anharmonic while the parameter $\Lambda = UN/(2J)$ is increased.
One can see that the value $z = 0$ is not accessible at any time if
\begin{equation}
 \frac{\Lambda}{2} z(0)^2 - \sqrt{1-z(0)^2}\cos(\phi(0))>1 \; .
 \label{self_smerzi}
\end{equation}
Under these conditions, one can identify a critical initial population imbalance
$z_c(0)$ at fixed $\Lambda$, and a critical $\Lambda_c$ at fixed $z(0)$:
for $z(0)>z_c(0)$ or for $\Lambda > \Lambda_c$ the particles are not any longer able to tunnel between the wells and $z(t)$ starts to oscillate around
a non-zero value, i.e., the condensate undergoes a self trapping. This self trapping is exhibited for the classical,
condensate limit, in which one can treat $\hat n_{a,b}$ as functions of time
in terms of their phases and numbers. When quantum fluctuations are considered,
the value $z=0$ can in fact be reached for large times
also for small values of $U/J$ \cite{Raghavan1999,Franzosi2000,smerzi2000,DellAnna}.

%
\subsection{Squeezing and Sensitivity}
\label{ses}
In this section, we introduce the core concepts and tools that will be used in our analysis on matter-wave interferometer sensitivity after reduction of quantum noise.
Interactions can lead to squeezed states; these are states in which the variance along one
axis is reduced at the cost of enhancing it along an orthogonal axis. Spin-squeezed states can be used to overcome the
shot-noise limit in interferometry \cite{Project,pezze}.  In fact, the Hamiltonian \eqref{H1} for the double-well system can be cast into the spin form
\begin{equation}
\hat{H}=-2J \hat J_x + \frac{U}{2}\hat J_z^2 + \delta \hat J_z \; ,
\label{H_S}
\end{equation}
after using the Schiwnger boson representation:
\begin{equation}
\begin{split}
& \hat{J}_x=\frac{1}{2}(\hat{a}^\dagger \hat b +\hat{b}^\dagger \hat a) \; ,\\
& \hat{J}_y=\frac{1}{2i}(\hat{a}^\dagger \hat b -\hat{b}^\dagger \hat a) \; ,\\
& \hat{J}_z=\frac{1}{2}(\hat{a}^\dagger \hat a -\hat{b}^\dagger \hat b) \; .
\end{split}
\label{Schwinger}
\end{equation}
One can thus see that the tunnelling term rotates a state on the Bloch sphere around the $x$-axis,
while the interaction term twists its components above and below the equator,
to the right and to the left respectively, while the twist rate increases with increasing the distance from the equator.

Different criteria have been introduced to assess whether a state is squeezed \cite{Gross_2012}. When dealing with $N$ particles, squeezing can be related to many-body entanglement, so that it turns out to be especially useful the criterion $\xi_S^2 \equiv N {(\Delta \hat J_{\vec{n}_3})^2}/(\langle \hat J_{\vec{n}_1}\rangle^2 + \langle \hat J_{\vec{n}_2}\rangle^2) < 1$, where $\vec n$ is a unitary vector and $\Delta \hat J_{\vec{n}_3}$ is the variance along the direction in which the state is squeezed. On the other hand, the sufficient condition for entanglement can be introduced
\cite{Entang_Squeez}:
\begin{equation}
 \chi^2 \equiv \frac{N}{F_Q[\hat \rho_{inp}, \hat J_{\vec{n}}]}< 1 \; ,
 \label{chi2}
\end{equation}
in terms of the quantum Fisher information 
$F_Q[\hat \rho_{inp}, \hat J_{\vec{n}}] = 4(\Delta \hat R)^2$, with $\hat R$ an Hermitian operator, solution
of $\{\hat R, \hat \rho_{inp}\} = i [\hat J_{\vec{n}}, \hat \rho_{inp}]$. Here, $\rho_{inp}= |\psi_{inp}\rangle \langle \psi_{inp}|$, $\hat R = \hat J_{\vec{n}}$ is the density matrix of the input state.
It is possible to show that $\chi^2 \leq \xi^2$, so that there are states which are entangled, $\chi^2<1$, but not
spin squeezed, $\xi^2 \geq 1$.

Quantum interferometry aims to resolve a phase shift $\phi$ below the shot-noise limit $\Delta \phi = 1/\sqrt{N}$. In general, the phase sensitivity is limited by the Quantum Cram\'{e}r-Rao (QCR) bound, which depends only on the input state:
\begin{equation}
 \Delta\phi_{QCR} = \frac{1}{\sqrt{F_Q[\hat \rho_{inp}, \hat J_{\vec n}]}} = \frac{\chi}{\sqrt{N}} \; ,
\end{equation}
so that the condition (\ref{chi2}) becomes also a necessary condition for measuring phase shift below the shot-noise limit:
$\chi < 1$ is required in order to have states there are usefully entangled for sub shot-noise sensitivity.

To estimate the phase $\phi$ accumulated after the interferometric sequence,
one measures an observable $\hat O$ which has $\phi$-dependent expectation values and variances.
The error propagation formula yields
\begin{equation}
\Delta \phi = \frac{\Delta \hat O}{\bigg| \partial \langle \hat O \rangle /\partial \phi \bigg|} = \frac{\xi_{S}}{\sqrt{N}} \; ,
\end{equation}
where $\Delta \hat O = \sqrt{\langle \hat O^2 \rangle - \langle \hat O \rangle ^2}$.
We can thus see that the sensitivity can be improved both by a larger slope of the expectation value of the operator as a function of $\phi$, or
by a smaller variance. In fact, the slope can be enhanced by the use of Schr\"odinger cat-type entangled states. However, these states are fragile against decoherence and, until now, have been realized only with few particles \cite{THL,Des}.
The variance of $\langle O \rangle$ instead, can be decreased using spin squeezing, whose aim is to reduce the projection noise.

In the following, we aim at finding the sensitivity with which the interferometer is able to measure a parameter coupled to $J_z$ in the Hamiltonian. During the phase accumulation stage, the spin rotates around the $J_z$ axis with frequency $\delta$, the second beam splitter converts the accumulated phase, $\phi = \int \delta dt$, into a measurable population imbalance between the two wells, and the sensitivity can be expressed as:
\begin{equation}
 \Delta \delta = \frac{\Delta \hat O}{\bigg| \partial \langle \hat O \rangle /\partial \delta \bigg|}  \; .
 \label{sensit}
\end{equation}
To this aim, we will mainly use two types of input states. One is a coherent state $|N,0\rangle$,
for which the sensitivity limit is given by the shot noise, i.e., $\Delta \delta = {1}/{\sqrt{N}}$. The other is is a Twin Fock state, which is a perfectly number-squeezed state
$|N/2,N/2\rangle$ (in the sense defined by $\chi_N$),
for which interferometry beyond the classical limit was studied \cite{twin}. Being $\Delta z = 0$,
the phase is completely indeterminate and  this state is represented on the Bloch sphere by an arbitrarily-thin line
around the equator. A rotation of this state in the absence of interactions, does not lead to a population imbalance,
so that other operators but $z$ have to be used in order to make a measurement. Possible choices are $\hat J_z^2$, or the parity operator $\Pi_b = e^{i\pi n_b}$ \cite{parity}.

\section{Inter-Well Interactions}
\label{inter-well section}
\begin{figure}[H]
\centering
\includegraphics[width = 0.55\textwidth]{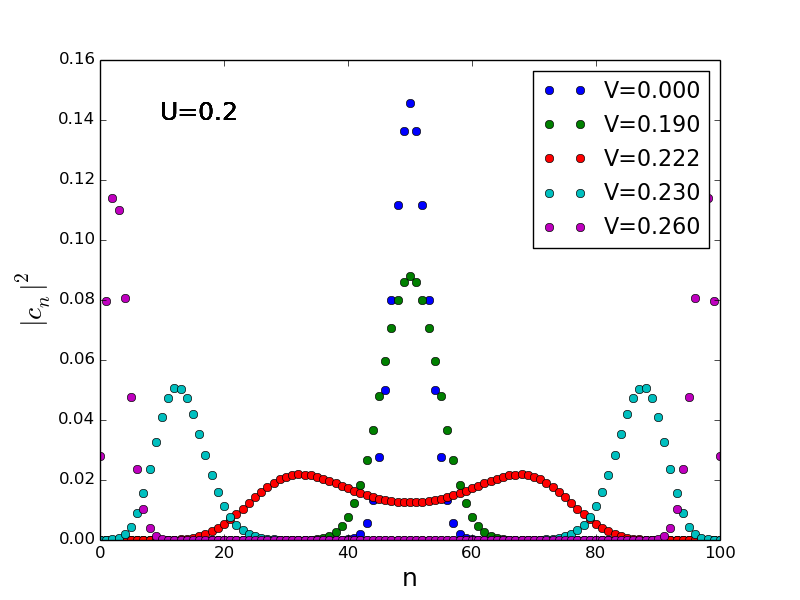}
\caption{Probabilities $|c_n|^2$ for the ground state of Hamiltonian (\ref{H}) at different values of $V$ as in the legend, and for $U=0.2$ and $\delta=0$. $U$ and $V$ are scaled in units of $J$.}
 \label{fig_2}
\end{figure}

We are now ready to explore the effect of inter-well interactions. The Hamiltonian for the 2M model now reads
\begin{equation}
 \hat H_{IW} = \hat H + V\hat n_a\hat n_b \; ,
 \label{H}
\end{equation}
where $\hat H$ is given in (\ref{H1}). In the case $V \neq 0$ and $J=0$,
it can be shown that, for $V>U$, Equation (\ref{H}) admits the NOON state as the ground state, even if the interactions between bosons are repulsive,
see Figure \ref{fig_2}.
This can be understood in terms of the interplay between inter- and intra-well interactions \cite{Mazzarella}: the former leads to a balanced population among the two wells, while the latter is minimum when one of the on-site average occupations vanishes.

After using the notation $\hat n=({\hat n_a - \hat n_b})/{2}$, Hamiltonian (\ref{H}) can be cast in the form 
\begin{equation}
\begin{split}
 \hat H & = -J(\hat a^\dagger \hat b +\hat b^\dagger \hat a) + \frac{U}{2}(\frac{4\hat n^2 + N^2}{2} - N ) + \frac{V}{4}(N^2- 4 \hat n^2) + \delta \hat n\\
 & = -J(\hat a^\dagger \hat b +\hat b^\dagger \hat a) + \frac{N^2}{4}( U+V) + \hat n^2( U-V)-\frac{U}{2} N  + \delta \hat n \; .
 \end{split}
 \label{inter_inter}
\end{equation}
We thus see  that when inter- and intra-well interactions are equal, the system behaves as if there were no interactions. On the other hand, for $V=2U$ the behavior is the same of a system with attractive interactions, i.e., with $-U$ coupling. As we shall see, this concept is crucial to our system, leading to the possibility of devicing a useful tool.

\section{Beam Splitter}
\label{BS}

We are now ready to analyze the behavior of
the first beam splitter acting in the interferometric process. We perform our study for two different initial states: a Twin Fock state $|N/2,N/2\rangle$ (Section \ref{TF_sec})
and a $|N,0\rangle$ state (Section \ref{N0_sec}). For the sake of comparison, we also study the sensitivity reached when a NOON state,
$({|N,0\rangle + |0,N\rangle})/{\sqrt{2}}$, is created after the beam splitter (Section \ref{NOON_sec}).

The general process used to realize a beam splitter for an atomic interferometer is to let the particles tunnel after
lowering the inter-well barrier in the case of a double-well interferometer, or by coupling the two states with resonant light, in the case of an internal two-level system interferometer.
In characterizing the beam splitter, we use the sensitivity as the full width at half maximum (FWHM)
of the narrowest peak in the phase probability distribution. The latter can be calculated as \cite{PS06}:
\begin{equation}
 \begin{split}
  P(\phi,t) & =  \frac{N+1}{2\pi}\langle \phi | \psi_{out}(t)\rangle \langle \psi_{out}(t)| \phi \rangle =\\
  & = \frac{1}{2\pi} \bigg|\sum_{n=0}^{N} c_n^{*}e^{i(N/2+n)\phi}\bigg|^{2} \; ,
 \end{split}
 \label{phase_distr}
\end{equation}
where 
\begin{equation}
 |\psi_{out} \rangle = \sum_{n=0}^{N} c_n(t)|n\rangle \; ,
\end{equation}
is the state after the first beam splitter. The coefficients $c_n(t)$ are given by $c_n(t) = \langle n|\psi_{out}(t) \rangle$, $\{|n\rangle\}$ with $n=[0,N]$ being the $N+1$ Fock basis vectors for the two-mode model: $|n\rangle = |n\rangle_a |N-n\rangle_b $. The $|\phi \rangle$ are the normalized phase states $|\phi \rangle = (N+1)^{-1}\sum_{m=0}^N e^{i\phi(N/2+m)}|m\rangle$.

\subsection{Initial State Twin Fock}\label{TF_sec}

When the initial state is a Twin Fock (TF) state $|N/2,N/2\rangle$, the optimal splitting time is defined as the time after which the main peak in the phase probability distribution is the narrowest \cite{PS06}.
In the non-interacting case, the $50/50$ beam splitter is represented by the unitary transformation
$|\Psi_{out}\rangle = e^{-i \frac{\pi}{2} \big( \frac{\hat a^\dagger \hat b + \hat b^\dagger \hat a}{2} \big)} |\Psi_{in}\rangle$
and the optimal splitting time is given by $T_{BS} = \pi/(4J)$. The latter corresponds to a $\pi/2$ Raman pulse,
which in the following we will refer to as $T_{\pi/2}$. As discussed in \cite{PS06},
the sensitivity is at the Heisenberg limit, i.e., $\Delta\phi = \alpha/N^\beta$ with $\beta \sim 1$: from our numerical simulation, we obtain $\beta = 1.001 \pm 0.004$ and $\alpha = 5.03 \pm 0.05$. 
When local interactions are present ($U \neq 0$, $V=0$), the phase probability distribution broadens and smaller values of $\beta$ are found after the beam splitter, with respect to the non-interacting case: from our numerical simulation we fit the parameters values $\alpha = 1.1 \pm 0.1$ and $\beta = 0.53 \pm 0.02$. Finally, we observe that the optimal splitting time decreases as the interaction energy is increased \cite{PS06}.

\begin{figure}[H]
\centering
\includegraphics[width =0.4\textwidth]{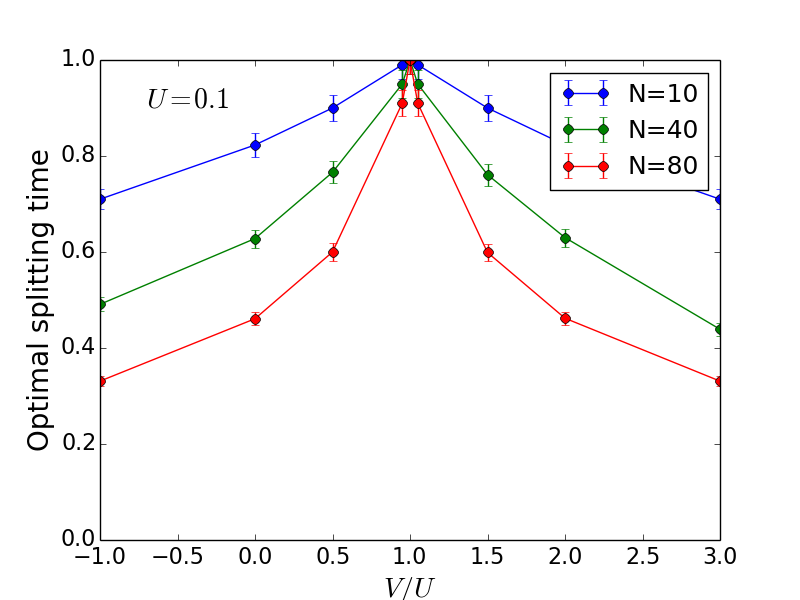}%
\includegraphics[width =0.4\textwidth]{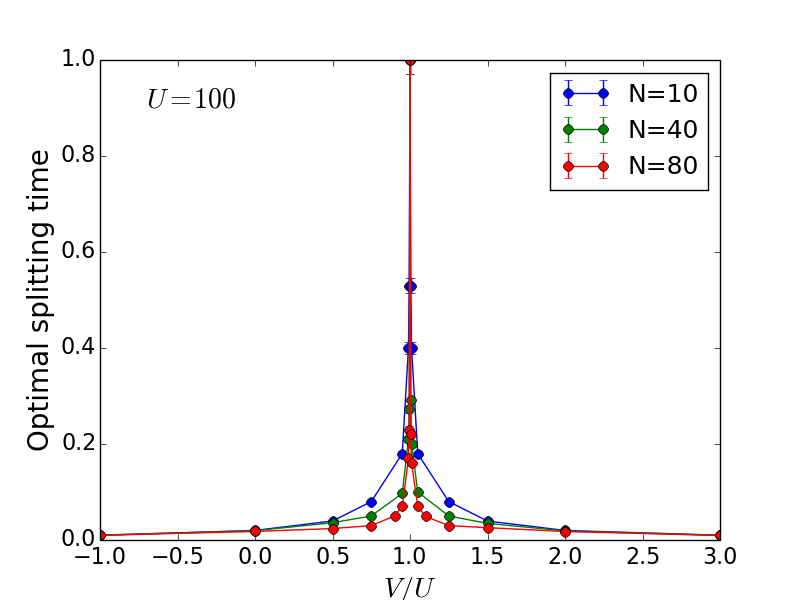}\\
\includegraphics[width =0.4\textwidth]{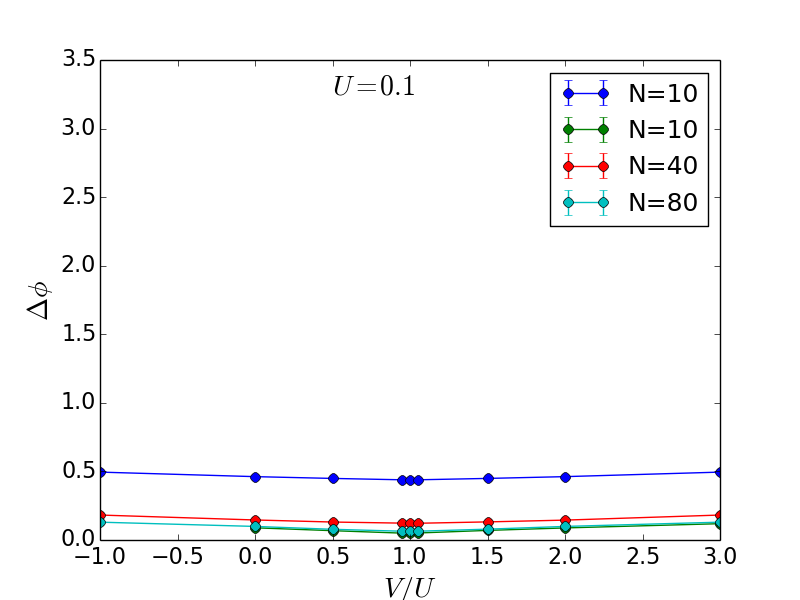}%
\includegraphics[width =0.4\textwidth]{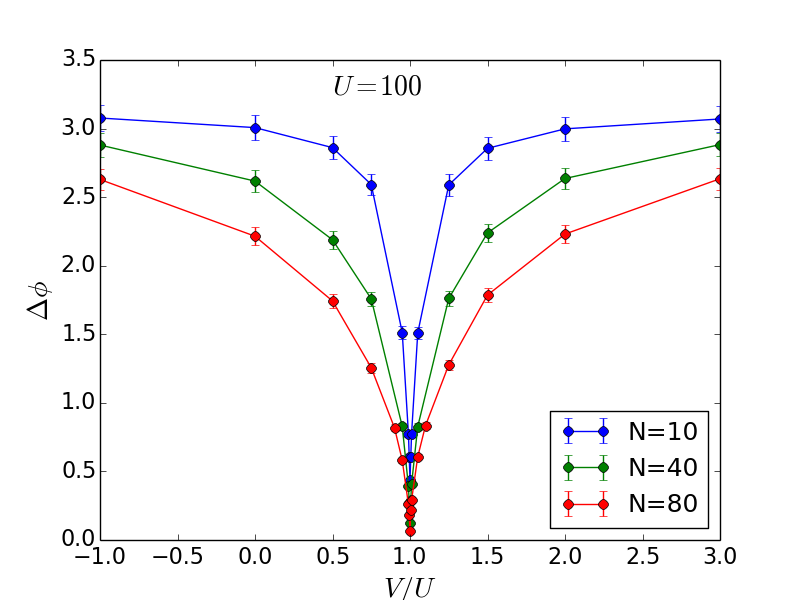}\\
\caption{Effect of inter-well interactions at the beam splitter for a Twin-Fock as input state. Optimal splitting time in units of $T_{\pi/2}$ (top) and $\Delta \phi$ (bottom) against $V$ for different values
  of particles number $N$ and interaction energy $U$ as in the legends. The value of $J$ is fixed to $J=0.5$.}
\label{fig_4}
\end{figure}

\begin{figure}[H]
\centering
\includegraphics[width = 0.4\textwidth]{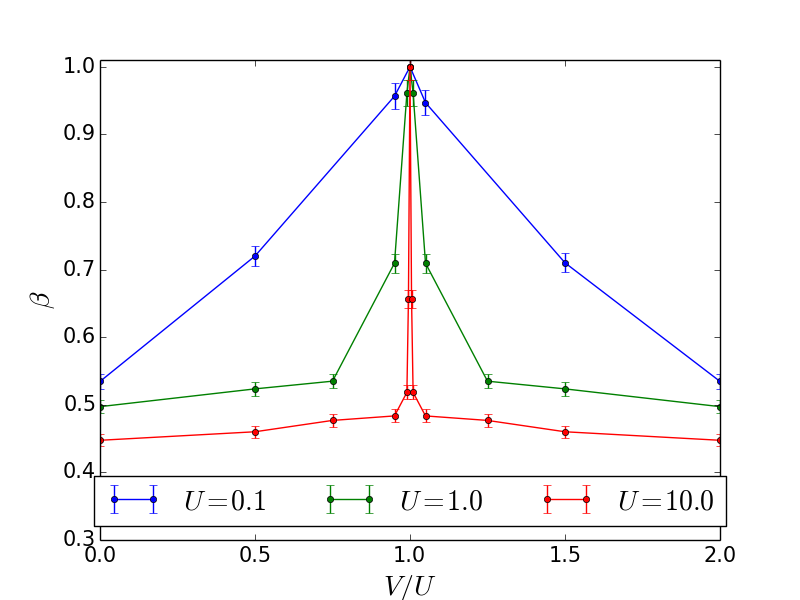}%
\caption{Effect of inter-well interactions at the beam splitter for a Twin-Fock as input state. Scaling parameter $\beta$ as a function of the inter-well interaction strength $V/U$ for different values of $U$.}
\label{fig_5}
\end{figure}

Let us now consider the effect of inter-well interactions. As already noticed,
in the limit $U/J \rightarrow -\infty$  the ground state of Hamiltonian (\ref{H1}) is a NOON state.
However, when the intra-well interactions are attractive, a collapse of the atomic cloud would take place, if the number of atoms exceeds a critical value \cite{Collapse}.
Varying the inter-well interactions far from the value $V=U$, the optimal splitting time decreases while the width
of the peak in the phase probability distribution increases, as illustrated in Figure \ref{fig_4}. The optimal splitting
time decreases with increasing $N$ at fixed $U$, and it is progressively less dependent on the number of particles while the intra-well interactions get larger. For $V=U$, the narrowest peak in the phase probability distribution
has the same width for each value of $U$. As predicted at the beginning of this section,
the scaling parameter $\beta$, defined from $\Delta \phi = {\alpha}/{N^\beta}$, is equal to that for the non-interacting case, in particular $\alpha =  5.03 \pm 0.05 $ and $\beta = 1.001 \pm 0.004$. In Figure \ref{fig_5}, the scaling parameter $\beta$ for different $U$ is reported against $V/U$: we notice that $\beta$ decreases with increasing the  interactions, except for $V=U$, where the non-interacting case is recovered.

\subsection{Initial State |N,0>}\label{N0_sec}
\begin{figure}[H]
\centering
\includegraphics[width = 0.4\textwidth]{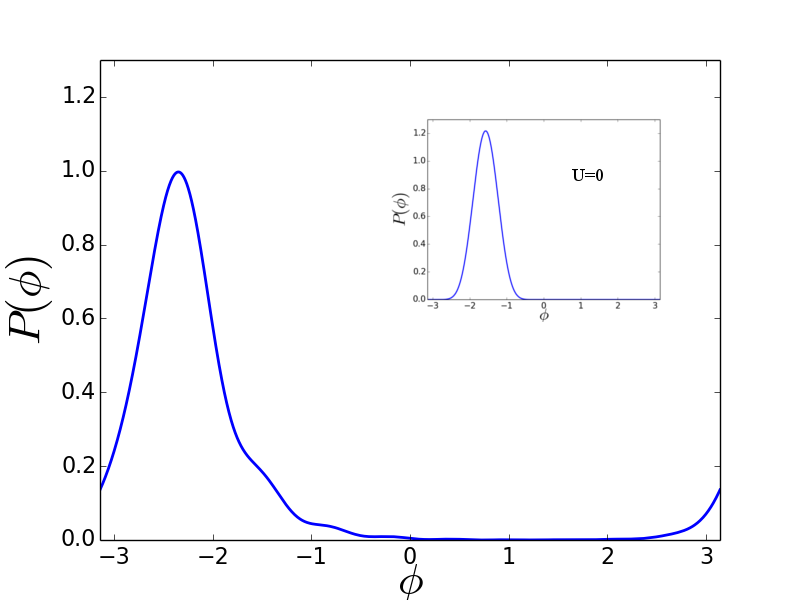}
\caption{Beam splitter for $|N,0\rangle$ as input state. Phase probability distributions after 
the first beam splitting for $U=0.3$, with $N=10$, $J=1$, $V=0$. Inset: the distributions for the non interacting case with $U=0$. }
\label{fig_6}
\end{figure}

We now consider the splitting process for the initial state $|N,0\rangle$, where the system is prepared with all the particles in one well. In this case, we define the optimal splitting time as the time at which the population is equally split among the two wells ($z = 0$).
We first analyze the non-interacting case. As for the Twin-Fock case as input state, the optimal splitting time is given by $T_{BS} = \pi/(4J)$, once again referred to as $T_{\pi/2}$.
For $U=V=0$, the shot-noise sensitivity is reached and $\Delta\phi = \alpha/N^\beta$, with $\beta \sim 0.5$.
When considering $U\neq0$ (with $V=0$), the condition ${UN}/{J} \leq 4$ must be fulfilled
in order to avoid self trapping.
Interactions are found to broaden the peak in the phase probability distribution, as shown in Figure \ref{fig_6}.  

\begin{figure}[H]
\centering
\includegraphics[width = 0.45\textwidth]{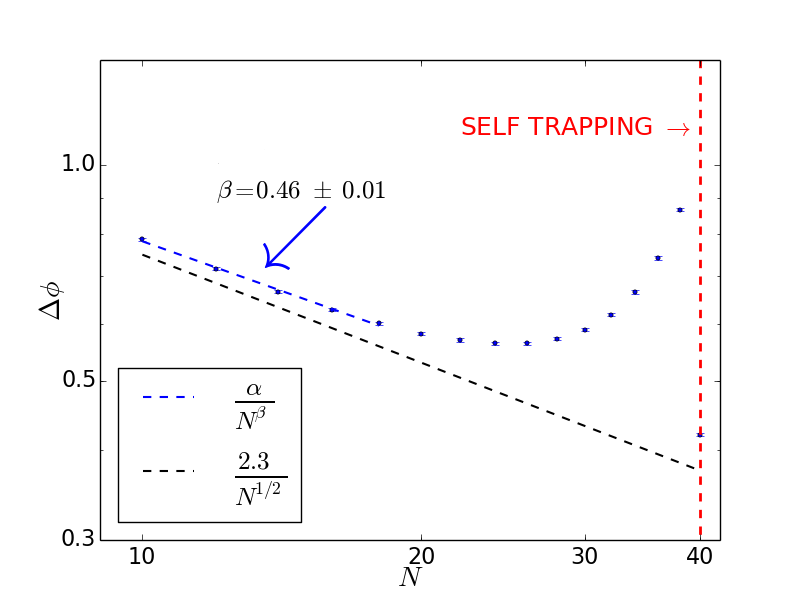}%
\includegraphics[width = 0.45\textwidth]{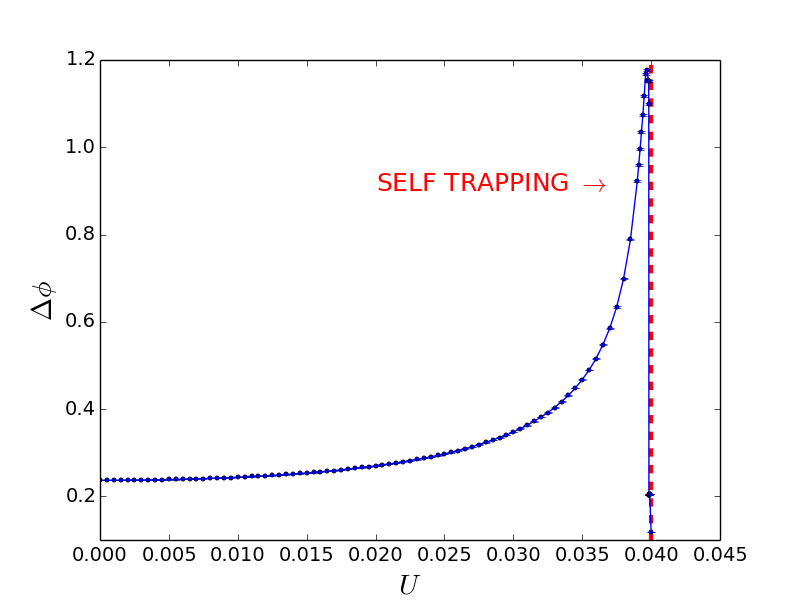}
\caption{Beam splitter for $|N,0\rangle$ as input state. $\Delta\phi$ plotted against $N$ at fixed $U=0.1$ (left) and against $U$ at fixed $N=100$ (right), with $V=0$. The fit parameters are found to be $\alpha = 2.3 \pm 0.1$ and $\beta = 0.46 \pm 0.01$.
The black dashed line represents the results for the non-interacting case, the red dashed line marks the self-trapping threshold.}
\label{fig_7}
\end{figure}

\begin{figure}[H]
\centering
\includegraphics[width = 0.45\textwidth]{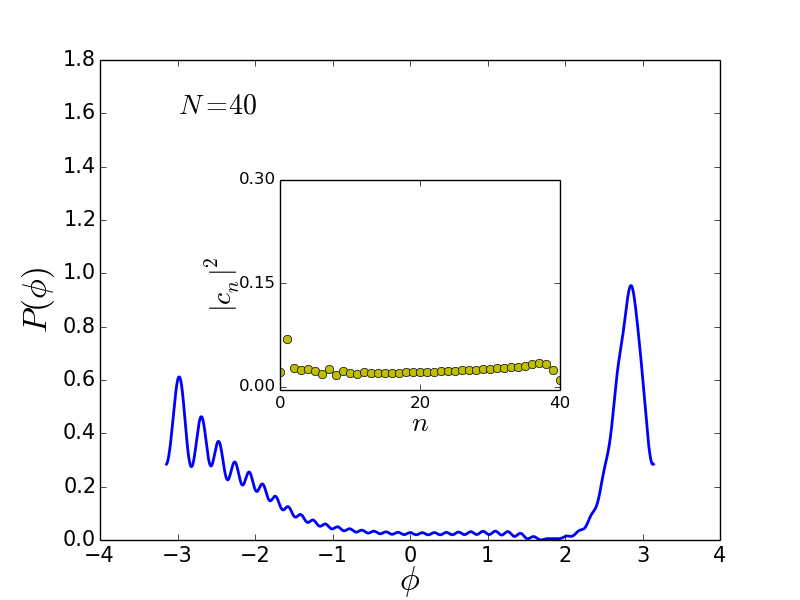}
\caption{Beam splitter for $|N,0\rangle$ as input state. Phase distributions (main figure) and relative-number distributions (inset) after the first beam splitter for $N=40$, self-trapping threshold. Here,  $U=0.1$ and $V=0$.}
\label{fig_8}
\end{figure}

After increasing the particles number, the width of the phase probability distribution peak decreases when $N$ is far from the self-trapping limit. Approaching the self-trapping limit, instead, the width begins to rise until
the self-trapping threshold is reached. As a result, a minimum is found at intermediate values of $N$, as visible in the left panel of Figure \ref{fig_7}. This can be understood by looking at the phase probability distributions for different values of the particles number.
By increasing $N$ near the self-trapping limit, the peak shifts and broadens. As displayed in Figure \ref{fig_8}, at the self-trapping threshold
the main peak splits in several sub-peaks, resulting in a narrower width. A similar behavior
is found when the self-trapping threshold is reached by increasing the value of the interaction energy at
fixed particles number. $\Delta \phi$ increases for energies lower than the self-trapping value and a minimum
is found at the self-trapping threshold. This is illustrated in the right panel of Figure \ref{fig_7}.

\begin{figure}[H]
\centering
\includegraphics[width = 0.45\textwidth]{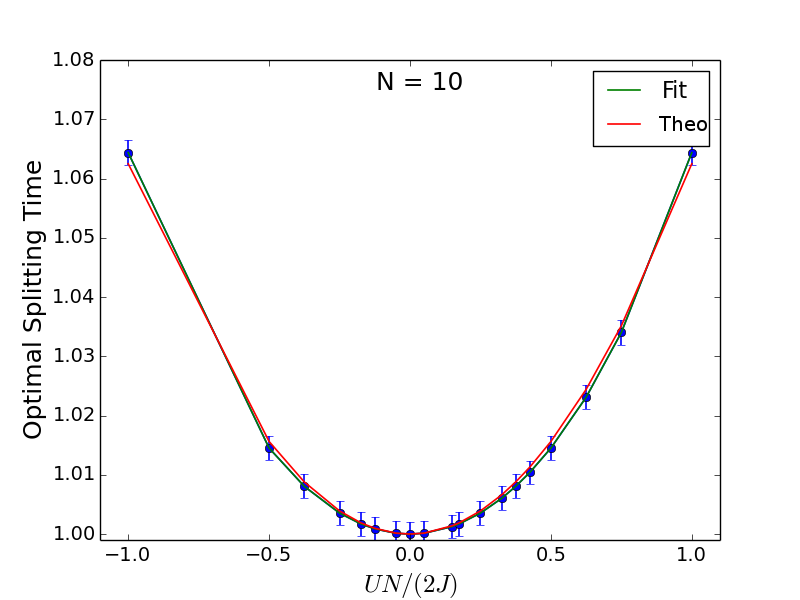}
\caption{Beam splitter for $|N,0\rangle$ as input state.  Optimal splitting time \textit{vs} $UN/(2J)$ obtained after varying $UN/(2J)$ at fixed $N=10$. Time is in units of $T_{{\pi}/{2}}(U=0) = {\pi}/({4J})$. The red curve corresponds to the theory and the green curve to the fit (see main text). Similar behavior is found by fixing $U$ and varying $N$.}
\label{fig_10}
\end{figure}

As to the optimal splitting time for the case of $|N,0\rangle$ as input state, this is found to increase with increasing $UN$, its values resulting in being the same for both repulsive and attractive
interactions. This behavior is visible in Figure \ref{fig_10}, where the optimal splitting time is displayed while varying $U$ at fixed $N$. Similar results are found by fixing $U$ and varying $N$. Now, the red curve in Figure \ref{fig_10} is given by the terms up to quadratic in the approximated equation for $t_{optimal}$:
\begin{equation}
 t_{optimal} \simeq \bigg[1 + \frac{1}{16} \Lambda^2 + \frac{9}{64}\frac{\Lambda^4}{16}+ \bigg(\frac{15}{48}\bigg)^2 \frac{\Lambda^6}{64}+...\bigg] \; ,
 \label{T_op}
\end{equation}
obtained in terms of $\Lambda \equiv {UN}/({2J})$ after integrating the equation for the population imbalance evolution
\begin{equation}
 \frac{\Lambda t}{2} = \int_{z(t)}^{z(0)} \frac{dz}{\sqrt{(\frac{2}{\Lambda})^2(1-z^2)-\big[ z^2 - \frac{2 H_0}{\Lambda}\big]^2}} \; ,
\end{equation}
between $z(0)=1$ and $z(t)=0$ and by scaling the time in units of $\pi/(4J)$. 
The slight discrepancy between fit and theory can be attributed to the small value of the particles number considered here.

\begin{figure}[H]
\centering
\includegraphics[width = 0.45\textwidth]{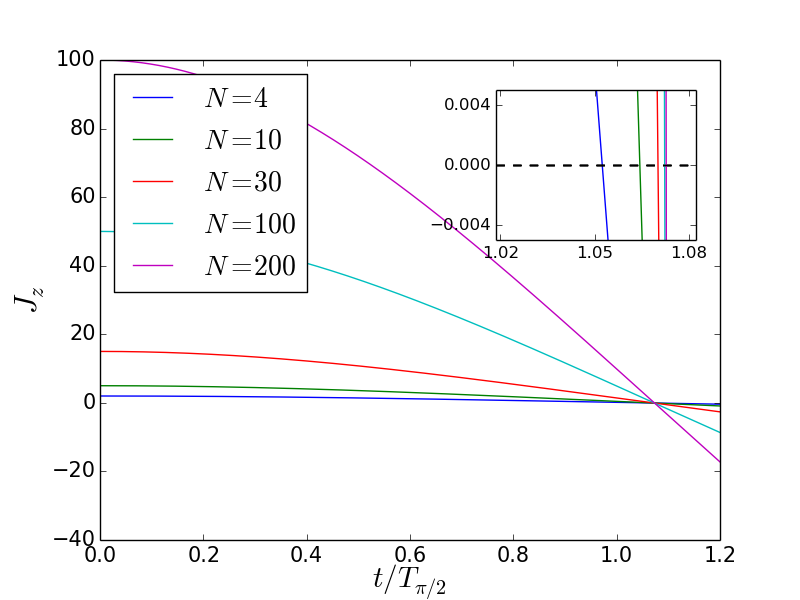}%
\includegraphics[width = 0.45\textwidth]{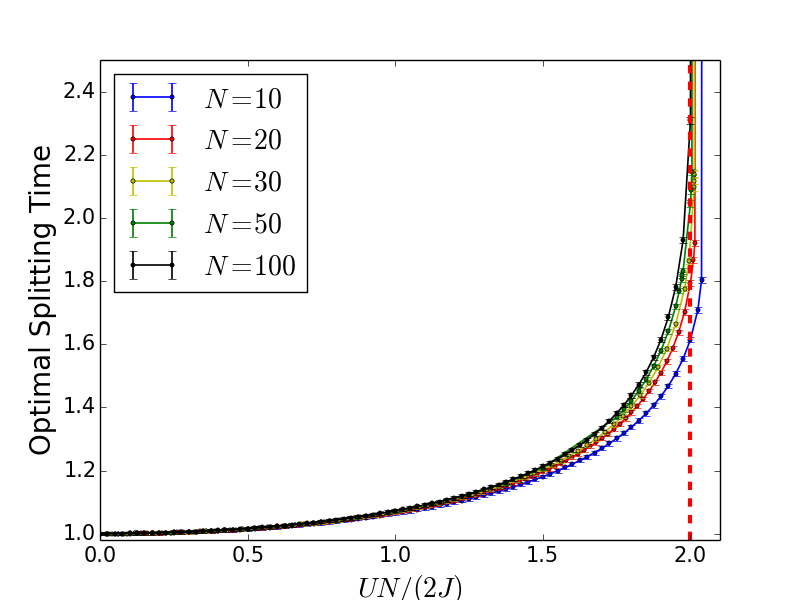}%
\caption{Beam splitter for $|N,0\rangle$ as input state. Search for universal behavior of the optimal splitting time.  ({\bf Left})  Time oscillations of the population imbalance $\hat J_z = (\hat n_a - \hat n_b)/2$ for the realization of the beam splitter, taken to be at the time for which $z=0$.
The values of the particles number $N$, as in the legend, and of the interaction energies $UN/(2J)$ are chosen so that $UN/(2J) = 1$. Inset: zoom near $J_z = 0$. While increasing the particles number, the difference between the corresponding optimal times decreases. ({\bf Right}) Optimal splitting time \textit{vs} $UN/(2J)$ for different values of $N$ as in the legend. The red dashed line marks the theoretical self- trapping threshold: for small $N$ this limit is higher than that predicted. Increasing the particles number, the behavior of the
optimal splitting time is dependent solely on $UN/(2J)$ and not on $N$ and $U$ separately. Time in units of $T_{\pi/2}$.}
\label{fig_11}
\end{figure}

Finally, looking for a universal dependence on $UN/(2J)$, the optimal splitting time has been found to be
dependent on the number of particles at fixed interaction energy (or viceversa) for small $N$ (large $U$), as shown in
Figure \ref{fig_11}. This behavior holds only for small values of the particles number, as we can infer from the inset
in the left panel of Figure \ref{fig_11}. By increasing the value of the particles number, the difference between the corresponding
optimal times decreases. In the right panel of Figure \ref{fig_11}, the optimal splitting time is plotted for different values of the particles number: we see that when $N$ is small, the self-trapping threshold is higher than that prescribed by Equation (\ref{self_smerzi}), valid for $N\gg1$. 

\begin{figure}[H]
\centering
\includegraphics[width = 0.44\textwidth]{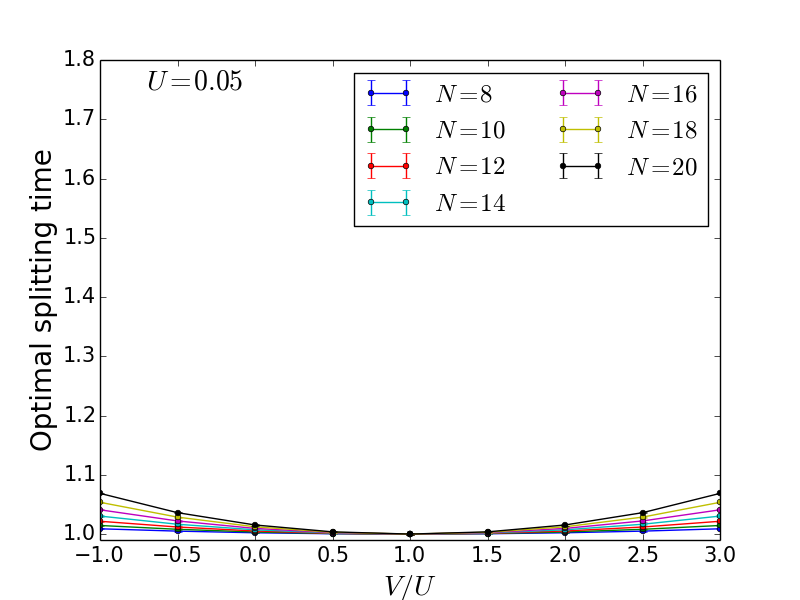}%
\includegraphics[width = 0.44\textwidth]{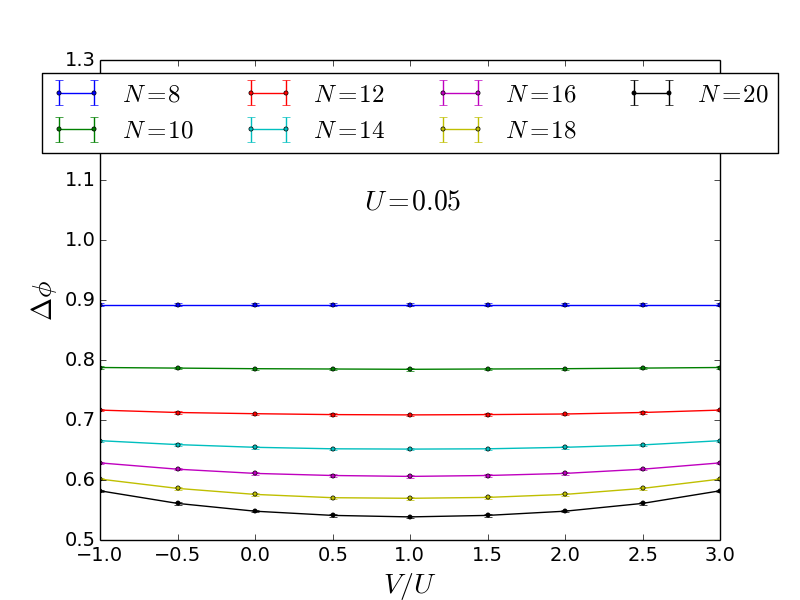}
\includegraphics[width = 0.44\textwidth]{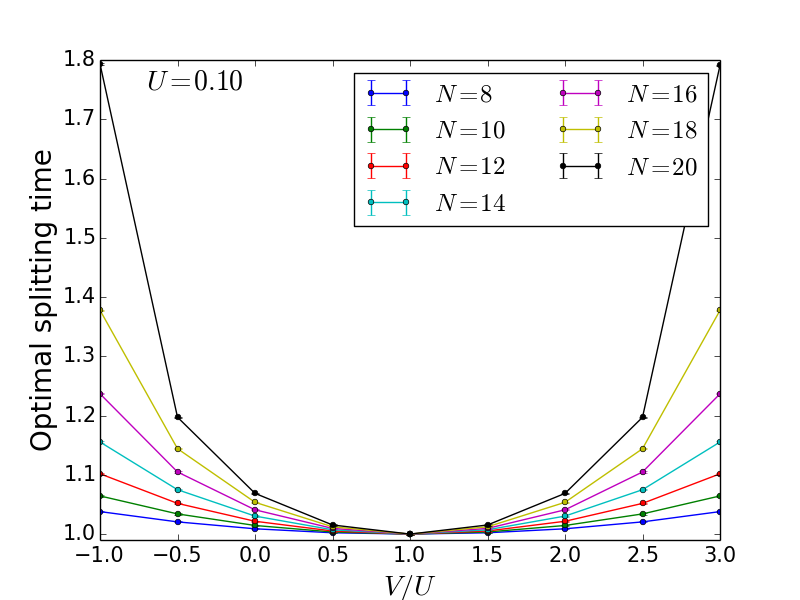}%
\includegraphics[width = 0.44\textwidth]{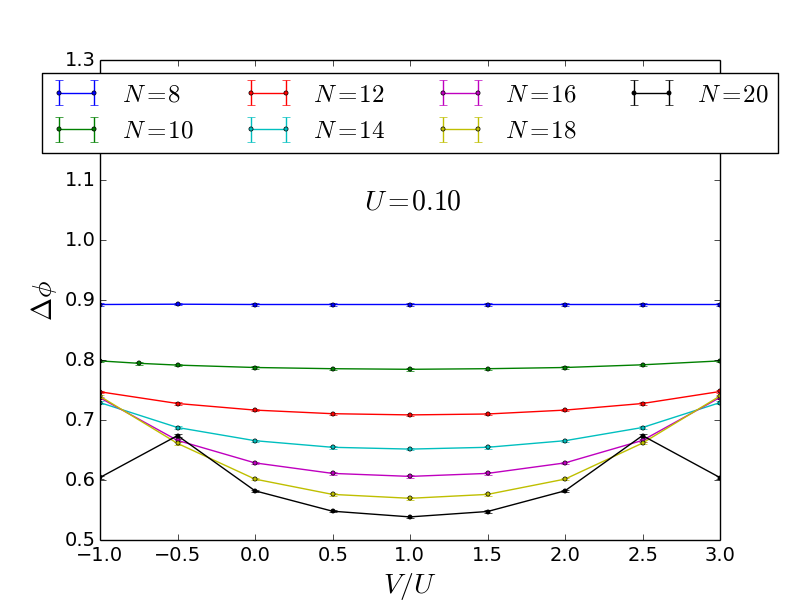}
\caption{Effect of inter-well interactions at the beam splitter for $|N,0\rangle$ as input state. Optimal splitting time in units of $T_{{\pi}/{2}}$ and FWHM of the peak
in the phase distribution as functions of $V/U$ for different values of $N$ as in the legends. ({\bf Top}): $U=0.05$. ({\bf Bottom}): $U=0.1$.}
\label{fig_12}
\end{figure}

We now turn to consider the effects of inter-well interactions. Varying the inter-well interactions,
the self-trapping limit can be reached for intra-well energy lower than the one satisfying Equation (\ref{self_smerzi})
($UN/(J)\leq 4$), as shown in the lower panel of Figure \ref{fig_12}. Indeed, we can see the self- trapping effect
in the curve for $N=20$: for $V/U = -0.5$ and  $V/U = 2.5$. The curve presents maxima with values higher
than the values found from the curves with $N=18$ and $N=16$. Notice that near the self-trapping limit, an increase in the
particles number leads to increasing values of $\Delta\phi$, as displayed in  Figure \ref{fig_7}. Local minima are found for
the values $V/U = -1$ and $V/U = 3$, corresponding to $U-V = \pm 0.2$ in the Hamiltonian (\ref{inter_inter}),
that is the self-trapping threshold.

\begin{figure}[H]
\centering
\includegraphics[width = 0.4\textwidth]{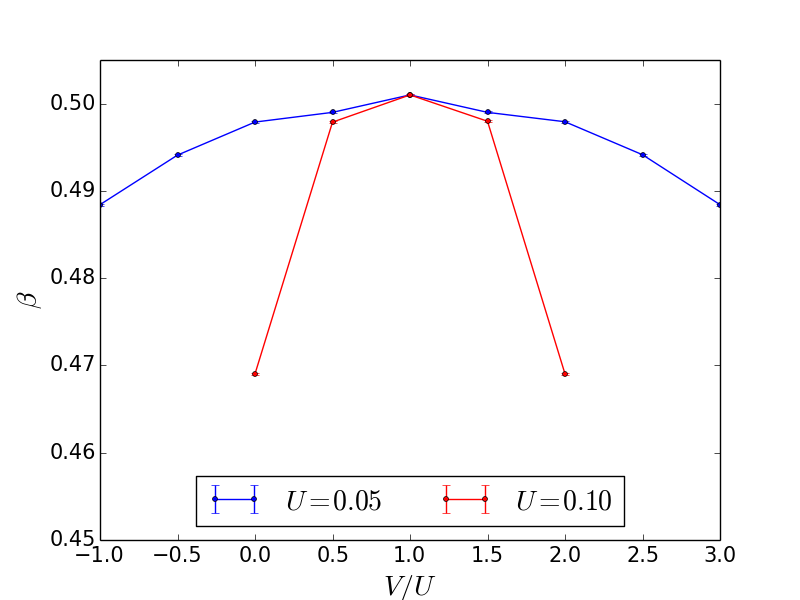}%
\includegraphics[width = 0.4\textwidth]{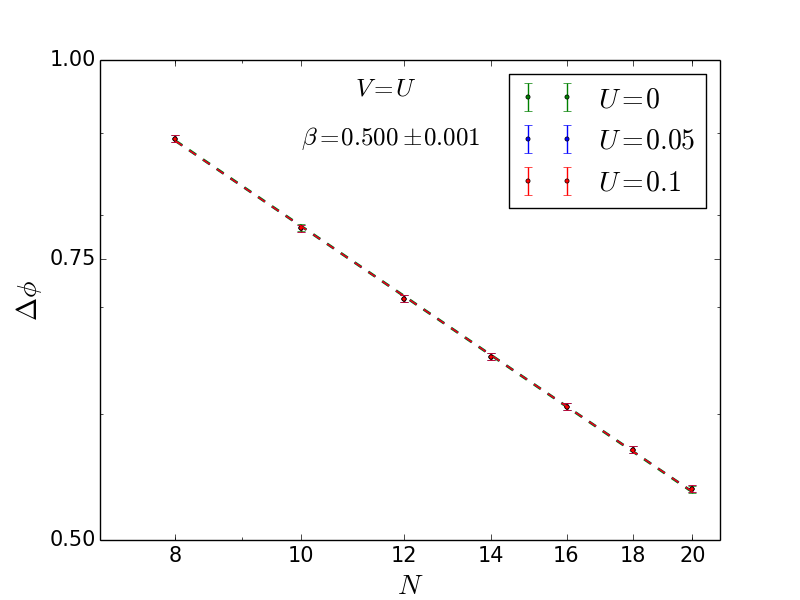}
\caption{Effect of inter-well interactions at the beam splitter for $|N,0\rangle$ as input state. ({\bf Left}). Scaling parameter $\beta$ \textit{vs} $V/U$ while varying the interaction energy. In the case with $U=0.10$ a fit for the value of $V/U$ near the self-trapping threshold would be meaningless. ({\bf Right})
$\Delta \phi$ \textit{vs} particles number for different energies and $V=U$: the same scaling
parameter $\beta = 0.500 \pm 0.001$ as in the non-interacting case is recovered ($\alpha = 2.37 \pm 0.01$).}
\label{fig_13}
\end{figure}

We close this analysis by noticing that the narrowest peak in the phase probability distribution for $V = U$ has the same width for each value of $U$. Similarly, the scaling parameter $\beta$ ($\Delta \phi = \alpha/N^\beta$ ) has the same value found in the non-interacting case, as reported in the right panel of Figure \ref{fig_13}. As it is evident from the left panel of Figure \ref{fig_13} displaying $\beta$ \textit{vs} $V /U$ for different $U$ values, increasing interactions worsen the scaling behavior, except when $V = U$, where the non-interacting case is recovered. 

\subsection{Discussion}\label{NOON_sec}
\begin{figure}[H]
\centering
\includegraphics[width = 0.55\textwidth]{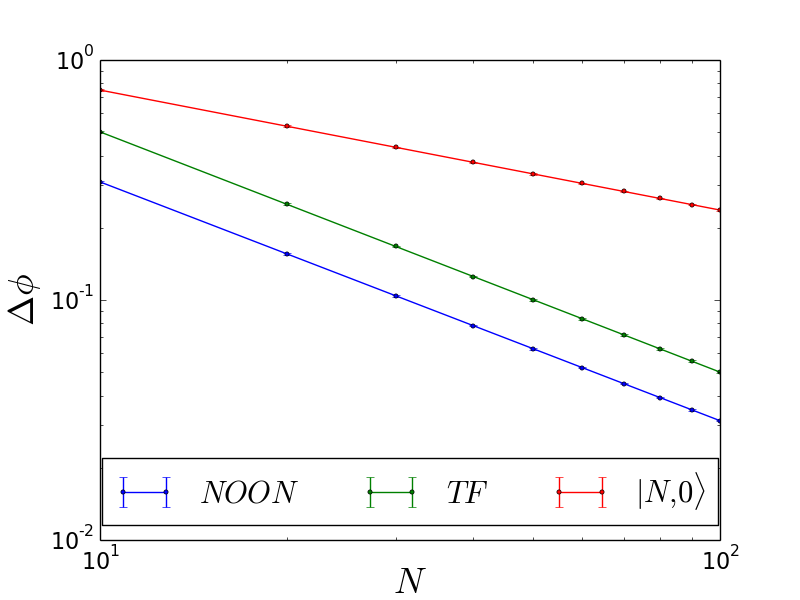}%
\caption{Sensitivity \textit{vs.} particles number $N$ in the non-interacting case for different input statea: $NOON$ (blue), $TF$ (green) and $|N,0\rangle$ (red).}
\label{fig_15}
\end{figure}

We conclude this section by discussing the comparison between the sensitivity that could be reached, if a NOON state 
$|NOON\rangle = {|N,0\rangle + |0,N\rangle}/{\sqrt{2}}$ be created after the beam splitter. Though creating a NOON state is a challenging task~\cite{pezze}, this is a useful analysis, since NOON states are known to provide a very good sensitivity in atom interferometry.
In particular, we find that the phase probability distribution for a NOON state presents equally spaced peaks, whose width depends on the particles number. The sensitivity is at a Heisenberg limit, with $\beta = 0.995 \pm 0.005$ and $\alpha = 3.07 \pm 0.04$. 
We are now in a position to compare the sensitivity that can be reached with the different states considered in this work. 
In Figure \ref{fig_15} we present the results for the non-interacting case.
Here, the sensitivity corresponding to the different analyzed states plotted against the particles number.
When the beam splitter is fed with the $|N,0\rangle$ state, the sensitivity scales as $\frac{\alpha}{\sqrt{N}}$,
close to the-shot noise limit. If, instead, a TF or NOON state is created after the beam splitter, $\Delta \phi = \frac{\alpha}{N}$ with the coefficient $\alpha_{NOON}< \alpha_{TF}$. The discussion of the non-interacting limit is useful when reference is considered to the possibility of adding inter-well
interactions. Indeed, as we discussed in this section, introducing inter-well interactions can be equivalent to be within non-interacting conditions. As a result, this work demonstrates that adding the non-local term may compensate this degradation and restore the best scaling and sensitivity.

\section{Conclusions}
\label{conclusions}

In this work, we  studied the effect of non-local interactions on the non-linear beam splitter in ultracold quantum
interferometers. After recalling known results on the two-mode (2M) model and the sensitivity of quantum interferometers,
we  analyzed the sensitivity for different input states. 
 We compared results obtained in the case of just local interactions with those found in the presence of non-local interactions, as those occurring in dipolar gases. In particular, we explored the sensitivity near the self-trapping limit and investigated the scaling behavior of the sensitivity as a function of the number of particles, while tuning the non-local interaction strength. Our analysis elucidates the role of inter-well interactions in compensating the sensitivity degradation induced by local interactions. This can be traced back to the fact that the inter-well coupling may effectively lead to an attractive interaction. This simple idea can be suitably exploited whenever intra-well interactions cannot be otherwise reduced. 
 
In the future, it would be interesting to implement our results to characterize the performance of a full interferometric scheme, in the presence of non-local interactions and/or when the interactions are turned on during the whole process.

{\em Acknowledgements:} Discussions with L. Pezz\'e and A. Smerzi are gratefully acknowledged. MLC acknowledges the support from the project PRA\_2018\_34 (``ANISE'') from the University of Pisa. GG work is supported by the Deutsche Forschungsgemeinschaft (DFG, German Research Foundation) under Germany's Excellence Strategy EXC 2181/1 - 390900948 (the Heidelberg STRUCTURES Excellence Cluster).MLC acknowledges the support from the project PRA\_2018\_34 ("ANISE") from
the University of Pisa.



\end{document}